CONDENSED MATTER PHYSICS

# Quasiparticle interference and quantum confinement in a correlated Rashba spin-split 2D electron liquid

Chi Ming Yim[1,2]*, Dibyashree Chakraborti[1,3], Luke C. Rhodes[1], Seunghyun Khim[3], Andrew P. Mackenzie[1,3], Peter Wahl[1]*

Exploiting inversion symmetry breaking (ISB) in systems with strong spin-orbit coupling promises control of spin through electric fields—crucial to achieve miniaturization in spintronic devices. Delivering on this promise requires a two-dimensional electron gas with a spin precession length shorter than the spin coherence length and a large spin splitting so that spin manipulation can be achieved over length scales of nanometers. Recently, the transition metal oxide terminations of delafossite oxides were found to exhibit a large Rashba spin splitting dominated by ISB. In this limit, the Fermi surface exhibits the same spin texture as for weak ISB, but the orbital texture is completely different, raising questions about the effect on quasiparticle scattering. We demonstrate that the spin-orbital selection rules relevant for conventional Rashba system are obeyed as true spin selection rules in this correlated electron liquid and determine its spin coherence length from quasiparticle interference imaging.

## INTRODUCTION

Spin-orbit coupling (SOC) in combination with inversion symmetry breaking (ISB) leads to the lifting of spin degeneracy (*1*, *2*) and promises spin manipulation without the need for magnetic fields. Possible applications of the Rashba effect include spin transistors enabling all-electronic spin manipulation (*3*) [for a recent review, see (*4*)]. Key parameters for a Rashba two-dimensional electron gas (2DEG) to be suitable for these applications include a sufficiently large Rashba spin splitting, as well as a spin relaxation length of the electrons that exceeds the spin precession length because of the spin splitting. The size of the Rashba spin splitting is directly linked to the spin precession length; a larger spin splitting will allow smaller structure sizes.

It is only recently that the surface states of delafossite oxides have been discovered, with potential functionalities for spin detection through surface ferromagnetism (*5*) and manipulation through the Rashba effect (*6*) in the same materials system. These materials offer an exceptional experimental platform for studying Rashba physics in a new limit where the ISB becomes the dominant energy scale. The delafossite metals (*7*–*9*) grow with astonishing crystalline purity (*10*) and show extremely long mean free paths as evidenced by bulk transport (*11*), transverse electron focusing (*12*), and even coherent Aharonov-Bohm–like oscillations (*13*). They have been shown to have surface states with very large Rashba splitting (∼70 meV and ∼150 meV in $PdCoO_2$ and $PdRhO_2$, respectively). This is seen on the transition metal oxide–terminated surfaces, and these splittings correspond to the full atomic SOC energies of Co and Rh, respectively. This is in strong contrast to what is usually seen in materials with large Rashba splitting, in which the observed splitting is a small fraction of the bare SOC energies of the relevant atoms (*14*–*16*). The Rashba physics seen in the delafossites arises because the unusual orientation of the transition metal octahedra in the delafossite structure leads to extremely large energy scales for the ISB (*6*). In the standard cases of large Rashba splitting, the observed effect is limited by the ISB energy scale, which is a weak perturbation compared to the atomic SOC energy scale. In the delafossites, it is not, so the full bare atomic SOC determines the splitting, leading to a giant spin splitting in a materials system composed of comparatively light elements.

Achieving huge spin splittings based on transition metals has obvious potential for spintronic applications (*17*), but the new situation comes with other characteristic features that are different to those in traditional giant Rashba systems, and merits further investigation. First, the different scale of ISB and SOC leads to a topologically different orbital and spin texture along the Fermi surface. The canonical case of a Rashba spin-split system with ISB much smaller than SOC exhibits orbital and spin textures that are directly coupled and exhibit the same chirality (compare Fig. 1A) (*18*), leading to a suppression of the same scattering vectors by the matrix element $\langle \Psi_k | \Psi_{k+q} \rangle$ through the orbital and spin channels. When the ISB dominates over the SOC, the two subbands exhibit the same orbital texture but opposite spin textures (Fig. 1B). This means that, only from the orbital texture, the dominant scattering vector for a Rashba system is expected to be suppressed, whereas the spin texture should result in the same dominant scattering vector as in a normal Rashba system.

Further, because the surface electronic structure is based on 3*d* (Co) or 4*d* (Rh) orbitals, the Rashba spin-split surface state of the delafossites is a strongly correlated 2D electron liquid (2DEL) rather than a simple 2DEG. The on-site repulsion, or Hubbard *U*, for Co is approximately 5 eV in $PdCoO_2$ (*11*) or $PtCoO_2$ (*19*), and the electronic effective masses in the Rashba surface states are typically more than 10 $m_e$ (*6*). This raises questions as to what the consequences are for the quasiparticle interference (QPI) in comparison to the previously studied systems in the limit where ISB is only a weak perturbation (*20*–*23*), how it is affected by electron interactions, and, particularly important for spintronic applications, how the spin coherence length is affected. To address these questions, we have investigated the quasiparticle scattering on cobalt oxide–terminated surfaces of $PdCoO_2$. We establish that the quasiparticle scattering is dominated by spin selection rules here, resulting in the same scattering pattern as would be expected from the spin-orbital selection rules for standard weak-coupling Rashba systems, and observe that the quasiparticles in the surface state have long spin coherence lengths that would, in principle, be suitable for spintronic applications.

[1]SUPA, School of Physics and Astronomy, University of St Andrews, North Haugh, St Andrews, Fife KY16 9SS, UK. [2]Tsung Dao Lee Institute and School of Physics and Astronomy, Shanghai Jiao Tong University, Shanghai 200240, China. [3]Max Planck Institute for Chemical Physics of Solids, Nöthnitzer Straße 40, 01187 Dresden, Germany.
*Corresponding author. Email: c.m.yim@sjtu.edu.cn (C.M.Y.); wahl@st-andrews.ac.uk (P.W.)

## RESULTS
### Identification of surface termination–surface polarity

$PdCoO_2$ consists of alternating layers of Pd and $CoO_2$ octahedra arranged in-plane in a hexagonal geometry, as illustrated in Fig. 2A. In the bulk, the Pd layers carry a charge of +1, while the $CoO_2$ layers carry a charge of −1 per unit cell. Because of the substantially stronger bonding within the $CoO_2$ octahedra than between the $CoO_2$ layer and the Pd, the cleaving is expected to occur between Pd and O, resulting in two possible clean surface terminations: a Pd-terminated surface and a $CoO_2$-terminated surface. Owing to the different charge on the layers, this leads to a polar surface that is either electron-doped (for Pd) or hole-doped (for $CoO_2$). Polarity-driven electronic reconstructions have been discussed extensively in the context of thin-film heterostructures of oxide perovskites, and the surface work function is an important parameter in understanding the role of the surface polarity as opposed to other mechanisms such as defect and vacancy formation (*24*, *25*).

Figure 2 (B and C) shows the topographic images obtained from the two surface terminations that exhibit large terraces but appear distinctly different. In addition, we frequently observe a disordered surface, which we attribute to a reconstruction of the Pd layer. The two surface terminations shown in Fig. 2 (B and C) show atomically flat surfaces; on the one shown in Fig. 2B we see atomic resolution with a lattice constant (~2.85 Å) consistent with the bulk crystal structure. Both exhibit standing wave patterns around defects due to QPI. Because of the polar nature of the surface terminations, we can assign the terminations by determining the local barrier height as a proxy for the work function. Figure 2D shows a measurement of the tunneling current $I(z)$ as a function of tip-sample distance $z$. From the slope, we obtain the local barrier height. We find that the surface shown in Fig. 2C exhibits a substantially lower local barrier height compared to the one in Fig. 2B. We can thus identify the surfaces shown in Fig. 2 (B and C) to be the $CoO_2$ and Pd terminations, respectively. We note that atomic resolution on the Pd termination is very difficult to observe, presumably because of the highly delocalized nature of the electronic states. The work function difference can also be determined independently from the image potential states in front of the surface, which are pinned to the vacuum energy,

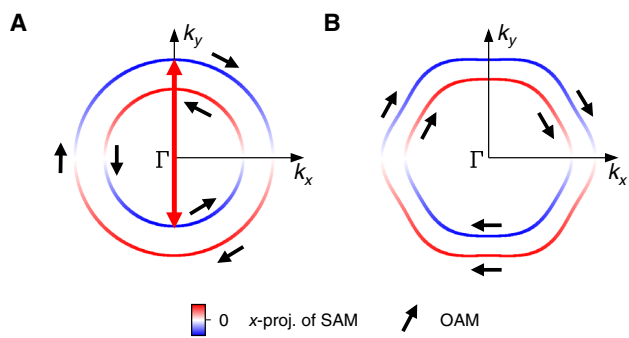

**Fig. 1. Orbital and spin textures of Rashba spin-split surface states in the limit of dominant SOC and dominant ISB.** (**A**) Schematic Fermi surface of a Rashba system where SOC dominates over the ISB. The spin and orbital textures are directly coupled and have the same chirality on the two subbands. The selection rules for QPI are the same for the spin and orbital component—both result in the same dominant scattering wave vectors. Black arrows show the orbital angular momentum (OAM), color encodes the x-projection of the spin momentum. (**B**) Schematic Fermi surface of the $CoO_2$-derived surface state where ISB dominates. Black arrows and colors as in (a). When ISB dominates over SOC, the orbital angular momentum (OAM) exhibits the same chirality on both bands, whereas the spin texture shows opposite chirality. This means that the selection rules for quasiparticle scattering lead to different wave vectors suppressed for the spin angular momentum (SAM) quantum numbers.

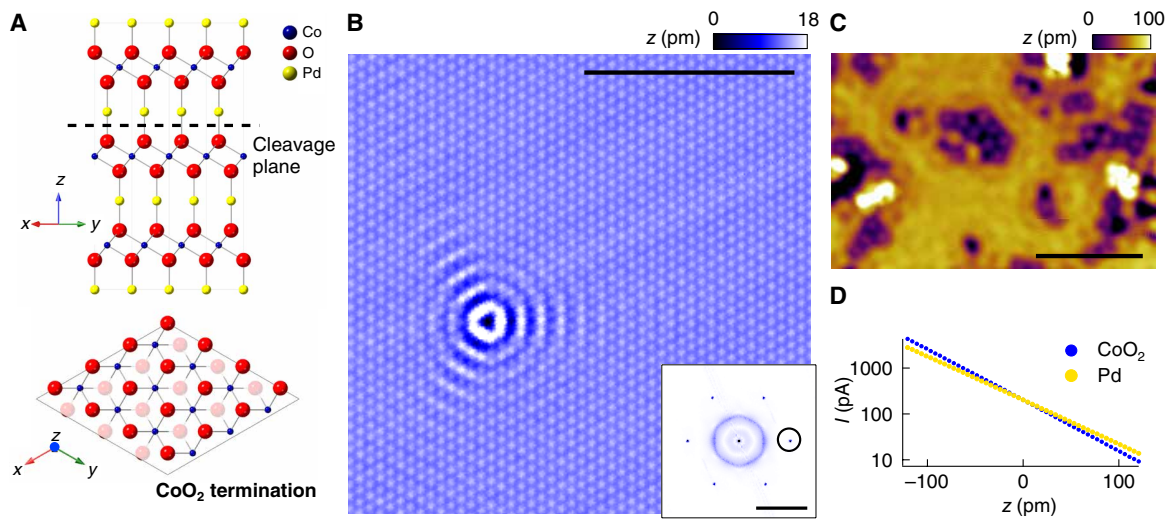

**Fig. 2. Surface terminations of $PdCoO_2$.** (**A**) Ball and stick model of the crystal structure of $PdCoO_2$. A dashed line marks the cleavage plane of this material. (**B**) Topographic STM image obtained from a $CoO_2$-terminated surface ($V = -50$ mV, $I = 50$ pA; scale bar, 5 nm). Inset, Fourier transformation of (B). Open circles mark the positions of the atomic peaks (scale bar, 2 Å$^{-1}$). (**C**) Topographic image of the Pd termination ($V = -80$ mV, $I = 50$ pA; scale bar, 5 nm). (**D**) Determination of the local barrier height from $I(z)$ measurements. $I$ versus $z$ plots taken from both the (blue) $CoO_2$-terminated and (yellow) Pd-terminated surfaces. $\Delta z = 0$ corresponds to the tip-sample distance, where the feedback loop was opened at a set point of $V = 100$ mV and $I = 200$ pA. Linear fits to the data reveal a difference of the local barrier height $\Delta\Phi = 3.2$ eV between the $CoO_2$ and Pd terminations (see also fig. S1).

yielding similar values for the local work function (see text S1 and fig. S1).

The large difference of the local work function between the two surface terminations of ~3.2 eV is a consequence of the polar surface termination. This difference in local barrier height would be consistent with a surface dipole because of a surface layer with only about half an electron charge and a permittivity on the order of 50, comparable to that found in other compounds with $CoO_2$ layers (26).

Having identified the two surface terminations, we will concentrate on the $CoO_2$-terminated surface and the Rashba 2DEL.

### Quasiparticle interference

To study the electronic structure across the Fermi energy, we have selected an atomically flat area of 38 nm$^2$ and acquired spectroscopic maps with more than 36,000 tunneling spectra consisting of 51 points each (see fig. S2 for the corresponding topographic image). In Fig. 3A, we show a layer of such a spectroscopic map of the normalized conductance $L(\mathbf{r}, V) = g(\mathbf{r}, V)/(\langle I(\mathbf{r}, V)\rangle_\mathbf{r}/V)$ for $V = 1$ mV, where $g(\mathbf{r}, V)$ is the differential conductance at position $\mathbf{r}$ and bias voltage $V$, and $\langle I(\mathbf{r}, V)\rangle_\mathbf{r}$ is the spatially averaged tunneling current at bias voltage $V$. The maps exhibit strongly dispersing QPI patterns because of the surface electronic states. In the real-space map in Fig. 3A, one can recognize well-defined hexagonally warped standing wave patterns emanating from the point defects on the surface. The long range of the standing wave patterns indicates a long phase and, hence, spin coherence length of the electronic states. In Fig. 3B, we show the corresponding momentum (q)-space map $\tilde{L}(\mathbf{q}, V)$ obtained by Fourier transformation and following symmetrization (see text S2 and fig. S3 for details). The q-space maps show a single hexagonally warped ring with a hole-like dispersion (additional layers of the map shown in fig. S4).

For a detailed comparison with the dispersion relation determined by angle-resolved photoemission spectroscopy (ARPES) in the occupied states, we use the tight binding model from (6) to perform T-matrix calculations of the QPI patterns. Figure 3C shows the Fermi surface of the $CoO_2$ termination obtained from the tight binding model. The Fermi surface consists of a pair of hole pockets centered at the Γ-point of the surface Brillouin zone, formed as a result of Rashba-like spin-orbit interactions. Naively, from the Fermi surface, one might expect quasiparticle scattering to take place via three different routes, that is, either between the two hole pockets or within each of the hole pockets. However, we observe only a single hexagon (Fig. 3B). This leaves interpocket scattering (indicated by solid red arrows in Fig. 3, B and C) as the only scattering pathway and demonstrates that spin is conserved throughout the scattering process. We note that this is different from the usual spin-orbital selection rule where the ISB is only a weak perturbation and the selection rule applies to the spin-orbital momentum $m_J$.

The quasiparticle pattern obtained from the T-matrix calculations is shown in Fig. 3D and shows excellent agreement with the experimental data.

### Analysis of the QPI data

For a detailed analysis of the QPI, we plot a cut of the $|\tilde{L}(\mathbf{q}, V)|$ map as a function of scattering momentum $\mathbf{q}$ and energy $eV$ along the high symmetry directions in Fig. 4A. We can see the dispersing scattering vector close to the zone boundary around the Fermi energy. Red dots in Fig. 4A show the allowed scattering momenta obtained from the T-matrix calculation. The corresponding $E - \mathbf{k}$ cuts along the high symmetry directions in the momentum space from the tight binding model are shown in Fig. 4B and fig. S5.

As for a conventional Rashba system, the spin splitting is expected to lead to a van Hove singularity in the density of states near the band edge, which should lead to a prominent feature in tunneling spectra [$g(V)$] (27). A tunneling spectrum obtained in a defect-free area is shown next to the experimentally determined dispersion in Fig. 4A, showing a main peak with a shoulder near the band top (see fig. S6 for a numerical fit of the two peaks). The density of states obtained from the tight binding model is shown for comparison in Fig. 4B. The calculated density of states shows a more pronounced double-peak structure near the band top, which indicates that the tight binding model may not capture the electronic structure close to the Γ-point as accurately as at lower energies.

For a detailed analysis of the QPI, we have determined the precise position and width in q-space of the QPI signal shown in Fig. 4A. The resulting dispersions along the Γ − M and Γ − K directions are shown in Fig. 4C. From the slope, we find effective masses at the Fermi energy along the Γ − M and Γ − K directions of −11 $m_e$ and −13.1 $m_e$, respectively, showing the correlated nature of the band. While these values cannot be directly compared with the band masses determined in ARPES (6), we can compare them to the average of the spin-orbit split bands, which is $-9\, m_e$ and $-12\, m_e$ in $\Gamma - M$ and $\Gamma - K$, respectively, slightly lower than the values we find.

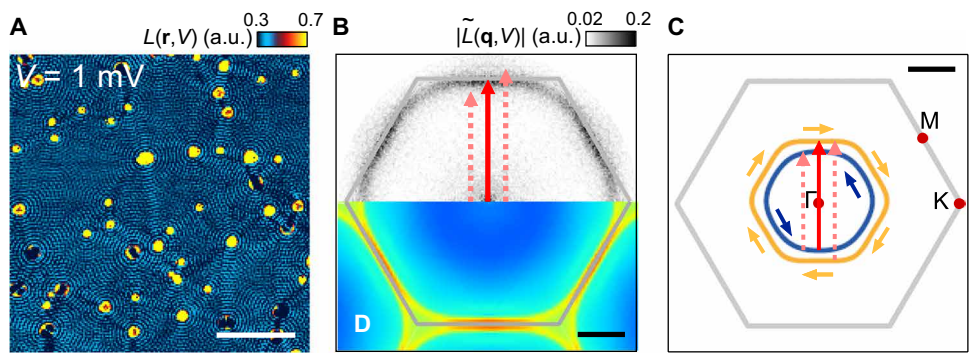

**Fig. 3. QPI of the $CoO_2$ termination.** (**A**) Real-space map of the normalized conductance $L(\mathbf{r}, V)$ measured from the $CoO_2$-terminated surface at a bias voltage of 1 mV ($V_s$ = 100 mV, $I_s$ = 400 pA, and $V_{mod}$ = 3 mV; scale bar, 10 nm). a.u., arbitrary units. (**B**) Corresponding momentum-space $\tilde{L}(\mathbf{q}, V)$ map (scale bar, 0.5 Å$^{-1}$). Hexagons mark the Brillouin zone (BZ) in the $k_z$ = 0 Å$^{-1}$ plane. A red arrow indicates the dominant scattering vector along the Γ → M direction. Pink arrows are the hypothesized scattering vectors assuming that intraband scattering also takes place. (**C**) Fermi surface of the $CoO_2$ termination. Grey lines mark the BZ boundary. Pockets of different colors have opposite spin texture in all directions, as a result of Rashba spin splitting. A red arrow marks the interband scattering vector as observed in the experiment; pink arrows mark the intraband scattering vectors not observed in the experiment. (**D**) Simulated QPI pattern at zero energy from the tight binding calculations, on the same scale as (B).

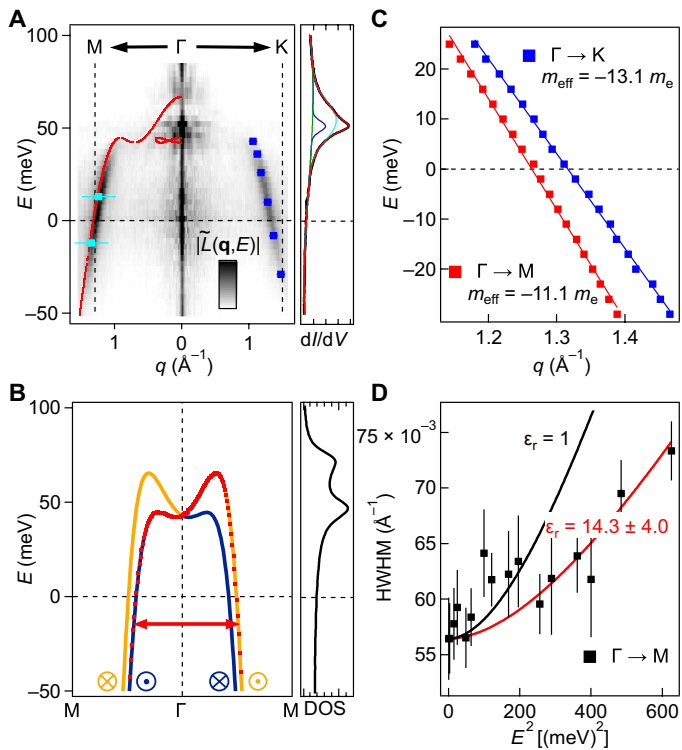

**Fig. 4. Dispersion relation from QPI.** (**A**) Cut of the $\tilde{L}(\mathbf{q},V)$ map in Fig. 3 as a function of momentum **q** and energy $eV$ along the high symmetry directions. Vertical dashed lines mark the **M** and **K** points of the BZ in the $k = 0$ Å$^{-1}$ plane. Red dots overlaid on the left are possible quasiparticle scattering vectors along the $\Gamma - M$ direction at different energies obtained from the band structure plotted in (B) assuming pure spin selection rules. Additional blue (cyan) data points overlaid on the right (left) are the scattering vectors found on a narrow terrace [width = 4.6 nm (=6.3 nm)] resulting from quantum confinement (see Fig. 5 and fig. S9 for more details). On the right, a d$I$/d$V$ spectrum taken in a defect-free area on the CoO$_2$ surface is shown, revealing the peak because of the Rashba-like surface states of CoO$_2$ centered at ~50 mV ($V_s$ = 100 mV, $I_s$ = 500 pA, and $V_{mod}$ = 0.25 mV). (**B**) Dispersion of the CoO$_2$-derived Rashba-like surface electronic states calculated from the tight binding model in (6). Bands of orange and blue colors represent the spin texture (see Fig. 3C). On the right, the corresponding density of states (DOS) is shown. (**C**) Dispersion of the quasiparticle scattering vectors along the $\Gamma - M$ (red) and $\Gamma - K$ (blue) directions. The effective mass along the two directions is estimated from their slopes. (**D**) Plot of HWHM of the scattering vectors along $\Gamma - M$ plotted versus $E^2$. Markers are experimental data. Solid lines show the expected behavior of the quasiparticle scattering lifetime (converted to momentum width) in 2D with a relative permittivity $\epsilon_r$ of 1 (black) and 14.3 (red), respectively (28).

The width of the quasiparticle scattering peak provides an estimate of the phase coherence length of the quasiparticles and, hence, of their lifetime. The three main contributions to the width of the QPI peak are a geometrical factor depending on the shape of the constant energy contours (and hence a property of the electronic structure of the material), disorder scattering, and electron-electron scattering. The width can be used to estimate the phase coherence length and thus a lower limit for the spin relaxation length. The width as a function of energy shows a minimum for quasiparticles right at the Fermi energy and increases with increasing energy. In Fig. 4D, we show the width [half width at half maximum (HWHM)] as a function of $E^2$ in the $\Gamma - M$ direction, where the signal of the QPI is strongest. The determined widths show an increase when moving away from the Fermi energy, thus confirming that the quasiparticles within the spin-split surface state bands behave like a Fermi liquid. We can compare the slope of the increase with theoretical predictions for the lifetime because of electron-electron scattering processes. In 2D, the line width $\Gamma$ in energy can be written as

$$\Gamma(\Delta) = \frac{\hbar}{\tau} = -\frac{E_F}{2}\left(\frac{\Delta}{E_F}\right)^2\left[\log\frac{\Delta}{E_F} - \frac{1}{2} - \log\frac{2q_{TF}^{(2)}}{k_F}\right] \quad (1)$$

where $E_F$ is the Fermi energy, $\Delta = E - E_F$ is the energy difference of the quasiparticle to $E_F$, $q_{TF}^{(2)} = \frac{2m^* e^2}{4\pi\epsilon\hbar^2}$ is the Thomas-Fermi wave vector in 2D, $m^*$ is the effective mass of electrons, $k_F$ is the Fermi wave vector, and $\epsilon = \epsilon_r\epsilon_0$ is the permittivity of the medium (28). For comparison with the experimental data, we convert $\Gamma(E)$ to $\Gamma_q(E) = \frac{1}{\frac{\partial}{\partial q}E(q)}\Gamma(E)$. We further account for the combined effects of the form factor of the QPI signal because of the shape of the constant energy contours and effects of disorder scattering through an additional broadening parameter $\Gamma_f$. This parameter is expected to be energy independent in the range investigated here because the Fermi surface retains its shape and disorder scattering is elastic. It can be determined from the width of the QPI signal for $E \to 0$ meV near the Fermi energy, where the only contribution from electron-electron interactions is due to thermally excited quasiparticles. The estimated Fermi temperature of our 2D surface electron liquid is 1700 K, so the electron-electron scattering is extremely weak at the Fermi energy at our measurement temperature of 4.2 K and can be neglected as a source of broadening, i.e., $\Gamma_q \to 0$. We account for energy-dependent broadening by adding it in quadrature to the broadening because of the form factor and disorder scattering; hence, we obtain the line width of the QPI signal, $\Gamma_{QPI}$, from $\Gamma_{QPI} = \sqrt{\Gamma_q^2 + \Gamma_f^2}$. In Fig. 4D, we plot the QPI linewidth $\Gamma_{QPI}(E)$ as a function of energy $E$ for a relative permittivity $\epsilon_r \approx 14$. That this value is lower than the one deduced from the work function difference, $\epsilon_r \sim 50$, is expected given that the wave functions of the electrons in the surface state extend into the vacuum and hence experience a lower permittivity than just the one of the CoO$_2$ layer. The minimal width of the QPI signal that we observe is $\Gamma_{QPI}(0) = 0.056$ Å$^{-1}$, which we can use to extract a lower limit for the phase coherence length $\chi$ of the electrons. The mean length scale over which the QPI patterns decay is $\xi_{QPI} = \frac{2\pi}{\Gamma_{QPI}} = 11.2$ nm, which corresponds to a phase coherence length of $\xi = 2\xi_q = 22.4$ nm (29). We note that while this value is extracted in the $\Gamma - M$ direction, we obtain a similar coherence length of 17.4 nm for the $\Gamma - K$ direction.

We stress that these values are lower estimates, because although the disorder contribution to $\Gamma_{QPI}$ results from a microscopic scattering process, the geometric one relating to the shape of the constant energy contours does not. It is also important to emphasize again that electron-electron scattering at the Fermi energy is completely negligible in this 2DEL at 4.2 K, because the experimental temperature is one-thousandth of the Fermi temperature. Under these conditions, we estimate the mean free path for electron-electron scattering to be tenths of a millimeter (see text S4), four orders of magnitude longer than our estimate for ξ. This is a crucial advantage of a fully metallic system with a large Fermi temperature compared with a doped semiconductor.

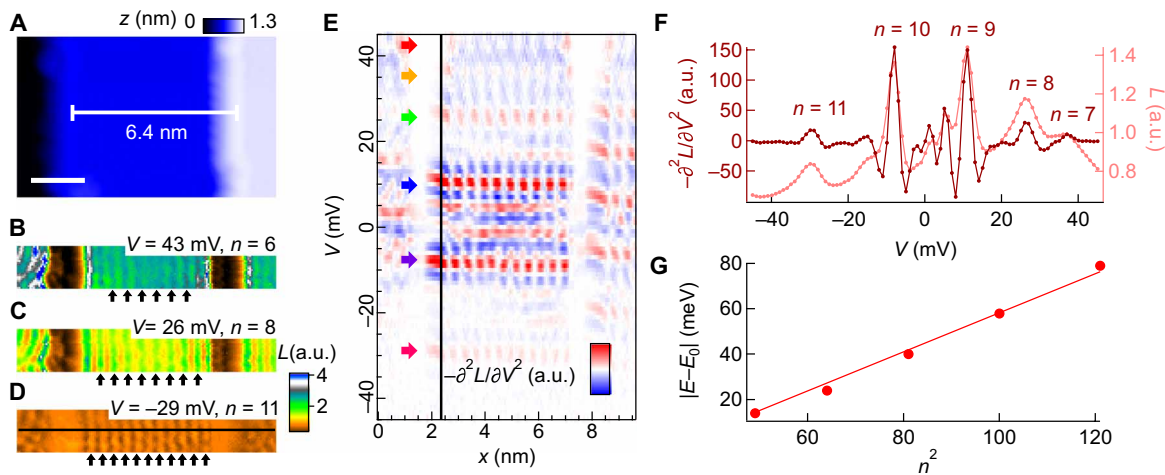

**Fig. 5. Spatial confinement of quasiparticle scattering on a narrow terrace.** (**A**) Topographic image taken from a stepped $CoO_2$-terminated surface. At the center of the image, the terrace is bound by a pair of step edges running along the [110] crystal direction and has a terrace of 6.4 nm ($V = -100$ mV, $I = 50$ pA; scale bar, 2 nm). (**B** to **D**) Maps of the normalized differential conductance $L(\mathbf{r}, V)$ across the region in (A) at three different bias voltages ($V_s = -45$ mV, $I_s = 400$ pA, $V_{mod} = 1$ mV). Arrows indicate the maxima in each of the standing wave patterns. (**E**) Negative curvature of the cross-section map, $-\partial^2 L(x, V)/\partial V^2$, extracted from the $L(\mathbf{r}, V)$ maps along the horizontal line marked in (D). Arrows indicate the bias voltage values where the standing wave patterns become most prominent. (**F**) Plot of $-\partial^2 L(x, V)/\partial V^2$ versus $V$ extracted at the location of the vertical line in (E). The corresponding $L(V)$ versus $V$ plot is also included. (**G**) Energy dependence of the resonator states plotted in the form of $|E_n - E_0|$ versus $n^2$, where $E_0$ denotes the energy of the band top, set at 50 meV. Markers are experimental data, and the solid line is a linear fit to the data, representing the $E_n \sim n^2$ behavior expected for a quantum well.

## Quantum confinement of Rashba split electronic states

While the QPI, because of scattering from a small number of impurities, does not provide direct information about the spin-split bands due to the spin selection rules, coherent superposition of higher-order scattering processes can lead to differences in the QPI patterns because of spin splitting (30). These higher-order scattering terms become important in resonator geometries, where quasiparticles scatter multiple times within a coherence length. Such a resonator geometry where the spin-split electronic states are confined between two step edges is shown in Fig. 5A. It consists of a $CoO_2$-terminated surface terrace with a width of 6.4 nm bound by a pair of step edges running along the [110] direction, leading to confinement in the Γ – K direction. In layers of the normalized conductance map $L(\mathbf{r}, V)$ taken from the same region (Fig. 5, B to D), one can see a series of standing wave patterns with different number of maxima and nodes. To further analyze the evolution of these patterns with energy, we show a linecut across the $L(\mathbf{r}, V)$ map plotted as a function of energy and position in Fig. 5E. The standing wave patterns appear at quantized energies $E_n$. The spacing between the quantized states increases with energy when moving away from the band edge as expected for a quantum well ($E \sim n^2$).

The quantized energy can more easily be seen in Fig. 5F, which shows a normalized spectrum $L(\mathbf{x}, V)$. Physically, $n$ is the main quantum number of the discrete energy states because of confinement. To compare the energies and characteristic length scales of these resonator states with the dispersion relation of the Rashba spin-split surface state, we plot the $\{E_n, q_n\}$ pairs obtained here onto the dispersion graph in Fig. 4A. All the data points fall exactly onto the dispersion relation arising from interband scattering between the spin-split bands. We therefore conclude that while we see clear quantum confinement of the spin-split electronic states, there is no sign of additional scattering processes between subbands with different spins. Similarly, comparing the resonator energies $E_n$ to the expected behavior for a quantum well, $E_n \sim n^2$, shows a consistent behavior in the range of energies where we observe clear resonator states (compare Fig. 5G). Close to the band edge, deviations become larger because the dispersion is not parabolic.

## DISCUSSION

We have been able to probe the spin selection rules for quasiparticle scattering in the unusual regime in which the ISB is the dominant energy scale. In these unusual circumstances, the spin-orbital selection rules for quasiparticle scattering, which were established in the opposite limit, become true spin selection rules.

Detailed analysis of the width of QPI features allows us to extract information about the phase coherence length of the quasiparticle states. Phase coherence implies, in particular, that the spin remains preserved over this length scale, and hence, this phase coherence length is effectively a spin coherence length. Our analysis reveals a behavior of the coherence length as a function of energy that is consistent with the expectation from 2D Fermi liquid theory if the permittivity of the surrounding medium is accounted for. Ab initio calculations of the closely related inelastic mean free path of surface states have previously been carried out for noble metals and systems with negligible electron correlations (31); the present system provides an opportunity to gauge the validity in the limit of a strongly correlated electron liquid.

Despite the strong correlations, the parameters of this Rashba split 2DEL are remarkable. The phase coherence length ξ ~ 22.4 nm, which provides a lower limit for the length over which the spin relaxes, is substantially longer than the spin precession length of approximately $L = 3.5$ nm obtained from the spin splitting [$L = \pi/\Delta k$ (3), where $\Delta k_{\Gamma-M} = 0.09$ Å$^{-1}$; see (6)]. This, in combination with the fact that spin remains a good quantum number because of the hierarchy of ISB and spin-orbit energies, means that the delafossite transition metal oxide surface state is in the regime required to realize spin manipulation on the length scales of current semiconductor technologies and without the need of heavy elements. Our observations were made on the surfaces of single crystals, but there seems to be no fundamental barrier to the eventual growth of equivalent quality delafossites in thin-film form. We believe that our observations provide strong motivation to do so.

In conclusion, we have demonstrated spin-selective QPI in the Rashba spin-split surface state of a correlated oxide and a 2DEL where the ISB becomes the dominant energy scale compared to the SOC. Our results are very well described by a $T$-matrix calculation of the scattering pattern, taking into account the spin selection rules. Detailed analysis of the line width of the QPI patterns allows us to extract information about the coherence length of the quasiparticles and compare it to theoretical predictions. The coherence length is substantially larger than the spin precession length, making this a promising system for spintronic applications. On a broader perspective, atomic-scale studies of the transition metal oxide terminations of other delafossites provide an opportunity to study the interplay between ISB, SOC, electronic correlations, and magnetic order (32).

## MATERIALS AND METHODS
### Sample growth
Single crystals of PdCoO$_2$ were grown in an evacuated quartz tube with a mixture of PdCl$_2$ and CoO by the following methathetical reaction: PdCl$_2$ + 2 CoO ⟶ 2PdCoO$_2$ + CoCl$_2$ (7, 33). The quartz tube was heated to 1000°C for 12 hours and held at 700 to 800°C for 5 days. To remove CoCl$_2$, the resultant product was washed with distilled water and ethanol. To obtain clean surfaces for scanning tunneling microscopy (STM) measurements, PdCoO$_2$ samples were cleaved in situ at ∼20 K in cryogenic vacuum.

### STM measurements
The STM experiments were performed using a homebuilt low-temperature STM, which operates at a base temperature of 1.8 K (34). Pt/Ir tips were used and conditioned by field emission with a gold single crystal. Differential conductance (d$I$/d$V$) maps and single-point spectra were obtained using a standard lock-in technique, with the frequency of the bias modulation set at 413 Hz. The results reported here were obtained at a sample temperature of 4.2 K.

### Tight binding model
In this report, we use the 22-orbital tight binding model of the CoO$_2$ layer of PtCoO$_2$ originally defined by Sunko et al. (6). To account for the quantitative differences between the PtCoO$_2$ and PdCoO$_2$ electronic structure, we reduce the octahedral crystal field splitting parameter, $C_O$, from 1.0 in (6) to 0.4 in this report. In addition, we apply a rigid chemical potential shift of $\Delta_\mu = -15$ meV. This has the effect of decreasing the $k_F$ values along the $\Gamma - M$ direction from 0.61 and 0.74 Å$^{-1}$ in (6) to 0.56 and 0.66 Å$^{-1}$, which is closer to the values obtained from ARPES measurements on PdCoO$_2$ (0.52 and 0.62 Å$^{-1}$) (6).

To model the QPI dispersion, we calculate the perturbation to the local density of states associated with a scalar nonmagnetic impurity using the $T$-matrix formalism

$$\delta N(\mathbf{q}, \omega) = -\frac{1}{2\pi i} \sum_{\mathbf{k}} Tr\left[\widehat{G}_0(\mathbf{k}, \omega) T(\omega) \widehat{G}_0(\mathbf{k} - \mathbf{q}, \omega) - \widehat{G}_0^*(\mathbf{k}, \omega) T^*(\omega) \widehat{G}_0^*(\mathbf{k} + \mathbf{q}, \omega)\right] \quad (2)$$

Here, $\widehat{G}_0(\mathbf{k}, \omega)$ describes the noninteracting Green's function at momentum $k$ and energy $\omega$

$$\widehat{G}_0(\mathbf{k}, \omega) = \frac{1}{(\omega + i\Gamma) - H(\mathbf{k})} \quad (3)$$

with $\Gamma$ defining the energy broadening of the calculation, which we set to 5 meV, and $H(\mathbf{k})$ describing the 22-orbital tight binding model of the CoO$_2$ layer of PtCoO$_2$ discussed above. Last, the $T$ matrix for a scalar nonmagnetic impurity is

$$\widehat{T}(\omega) = \frac{V}{\widehat{1} - V\sum_{\mathbf{k}} \widehat{G}_0(\mathbf{k}, \omega)} \quad (4)$$

In this report, we set $V = -100$ meV. It was found that increasing $V$ did not qualitatively change the calculated QPI patterns.

## SUPPLEMENTARY MATERIALS
Supplementary material for this article is available at http://advances.sciencemag.org/cgi/content/full/7/15/eabd7361/DC1

**Acknowledgments:** We acknowledge discussions with S. Bahramy, P. King, F. Mazzola, V. Sunko, and M. Watson. **Funding:** C.M.Y. and P.W. acknowledge support from the Engineering and Physical Sciences Research Council (EP/S005005/1), D.C. from the International Max Planck Research School for the Chemistry and Physics of Quantum Materials, L.C.R. from the Royal Commission for the Exhibition 1851, and A.P.M. from the Max Planck Society for the Advancement of Science. C.M.Y. acknowledges additional support from a Shanghai talent program. **Author contributions:** P.W. and A.P.M. conceived the project. C.M.Y. and D.C. performed STM experiments and analyzed the data. L.C.R. performed tight binding and $T$-matrix calculations. S.K. grew the crystals. C.M.Y., P.W., and A.P.M. wrote the manuscript. All authors discussed and contributed to the manuscript. **Competing interests:** The authors declare that they have no competing interests. **Data and materials availability:** All data needed to evaluate the conclusions in the paper are present in the paper and/or the Supplementary Materials. Additional data related to this paper may be requested from the authors. The data underpinning the findings of this study are available at https://doi.org/10.17630/024e3c43-44fb-469d-8f13-1ea82d159d47.

Submitted 8 July 2020
Accepted 22 February 2021
Published 9 April 2021
10.1126/sciadv.abd7361

**Citation:** C. M. Yim, D. Chakraborti, L. C. Rhodes, S. Khim, A. P. Mackenzie, P. Wahl, Quasiparticle interference and quantum confinement in a correlated Rashba spin-split 2D electron liquid. *Sci. Adv.* **7**, eabd7361 (2021).